\documentstyle[PASJadd,epsfig]{PASJ95}

\begin{document}
\setcounter{page}{1}

\title{Temporal $1/f^\alpha$ Fluctuations from Fractal Magnetic Fields\break
in Black Hole Accretion Flow}

\author{ Toshihiro {\sc Kawaguchi}\thanks{Research Fellow of the Japan
Society for the Promotion of Science} \ and Shin {\sc Mineshige} \\
{\it Department of Astronomy, Graduate School of Science, Kyoto
University, Sakyo-ku, Kyoto 606-8502} \\
{\it  E-mail(TK): kawaguti@kusastro.kyoto-u.ac.jp} \\
Mami {\sc Machida} and Ryoji {\sc Matsumoto} \\
{\it Department of Physics, Chiba University, Inage-ku, Chiba 263-8522} \\
and \\
Kazunari {\sc Shibata} \\
{\it Kwasan Observatory, Kyoto University, Yamashina-ku, Kyoto 607-8471}}

\abst{
Rapid fluctuation with a frequency dependence of $1/f^{\alpha}$
(with $\alpha \!\simeq\! 1\!-\!2$)
is characteristic of radiation from black-hole objects.
Its origin remains poorly understood.
We examine 
the three-dimensional magnetohydrodynamical (MHD) simulation data, 
finding that a magnetized accretion disk exhibits 
both $1/f^\alpha$ fluctuation 
(with $\alpha \!\simeq\! 2$) and a fractal magnetic structure (with the
fractal dimension of $D\!\sim\! 1.9$).
The fractal field configuration leads
reconnection events with a variety of released energy and of duration,
thereby producing $1/f^\alpha$ fluctuations.
}

\kword{Accretion, accretion disks --- Advection-dominated flow --- 
Black holes --- Fractal --- Magnetohydrodynamics}

\maketitle

\section{Introduction}

Apparently random temporal fluctuations from 
Galactic black-hole candidates (BHCs; van der Klis 1995)
and from active galactic nuclei (AGNs; Ulrich, Maraschi \& Urry 1997) 
have led many astronomers to recognizing how complex 
the nature behaves.
The light curves are neither periodic nor random around some mean.
Rather, they are seemingly composed of
shot events with a variety of peak intensities and durations 
(Negoro et al.\ 1995).
Number of analyses of X-ray light curves 
and optical AGN light curves show that 
power spectral density (PSD) is flat at lower frequencies ($f$) 
and
is power-law ($\propto f^{-\alpha}$ with $\alpha \simeq 1 - 2$) 
at higher frequencies.
The break frequency corresponds to a few second for BHCs and to a few
years for AGNs.
More, $1/f^{\alpha}$ fluctuations are ubiquitous in natural behavior, 
although their origins have been unsolved 
(Tajima \& Shibata 1997; Cable \& Tajima 1997).
The significance of the $1/f^{\alpha}$ noise is that
it contains a long-term memory (Press 1978).
It has been a puzzle how $1/f^{\alpha}$ fluctuations can  
arise in black-hole accretion flows (or disks) under realistic circumstances.

Among number of suggestions for a possible mechanism of variability,
the most promising one is magnetic flares 
(Wheeler 1977; Galeev, Rosner \& Vaiana 1979).
It has been established through the X-ray observations
by the Yohkoh satellite that the solar flares are triggered by
magnetic reconnection (Shibata 1996).
In fact, solar soft X-ray variation exhibits $1/f^\alpha$ 
fluctuations (UeNo et al.\ 1997).
Similarly, sporadic magnetic reconnection events which can occur 
in accretion flow may be responsible for the variability 
of black-hole objects, as well (Mineshige, Kusunose \& Matsumoto 1995). 

To prove this conjecture, we examine the three-dimensional data
of global, MHD disk calculations first made by 
Machida, Hayashi \& Matsumoto (1999).
They calculated how magnetic field evolves 
in a rotating disk initially threaded by toroidal ($B_\varphi$) fields.
Since no cooling is taken into account in computations, the simulated
disk is advection-dominated (Kato, Fukue \& Mineshige 1998),
rather than radiation-dominated as in the standard disk.
Then, the system we analyze corresponds to BHCs in 
the hard (low) state, in which fluctuations are largely enhanced and
whose spectra can well be reproduced by advection-dominated flow
(Narayan, McClintock \& Yi 1996).
Magnetic fields are amplified with time via 
a number of MHD instabilities together with differential rotation.  
The maximum field strength is determined 
either by field dissipation by reconnection 
or field escape from accretion flow via Parker instability.  
As a result, the mean plasma $\beta$, the ratio of gas pressure 
to magnetic pressure, finally reaches $\sim$10 
irrespective of initial values of $\beta$.
Locally, however,
even low-$\beta$ ($<1$) regions appear;
inhomogeneous structure arises consequentially [see also similar
discussion by Abramowicz et al. (1992) but for non-magnetic cases].

In the present paper, we thus analyze temporal and spatial behavior of
the accretion flow in order to clarify the origin of $1/f^\alpha$ 
fluctuations in black-hole objects.
Results of the analysis are presented in the following section. 
The final section is devoted to discussions. 

\section{Temporal and Spatial Analysis of Accretion Flow}

Figure 1 displays the light curves ({\rm A}) and their PSDs ({\rm B})
of the simulated disk obtained 
in the quasi-stationary state.
Here, we assume that magnetic reconnection events contribute
much to the energy output and thus calculated time variation
of $\int \eta j^2 dV$ (with $\eta$ and $j$ being the electric resistivity 
and electric current density) integrated over almost whole disk.
High frequency sides of the PSDs show nearly
power-law decline with an index of $\alpha \sim 2$, in agreement with
the observations.
This is, in a sense, amazing that without any fine-tuning of parameters
or special assumptions an MHD disk naturally 
gives rise to $1/f^\alpha$ fluctuations over three orders of frequency ranges.
On the other hand, low frequency sides flatten at frequencies smaller
than a reciprocal of
several rotation timescale at a reference radius, 
which is also consistent with the observations 
(van der Klis 1995; Ulrich et al.\ 1997).
\begin{figure}[t]
\hbox{\psfig{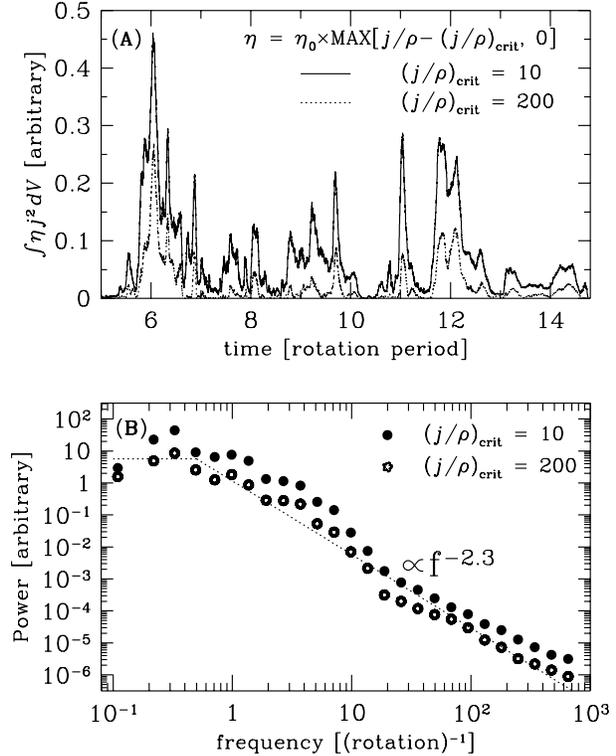}}
\caption
{
({\bf A})
Typical light curves of the simulated MHD disk.  
Here, we assume that radiation is
predominantly due to field dissipation by magnetic reconnection,
thus plotting the temporal variation of the $\eta j^2$ 
integrated over almost whole disk.
Electric resistivity $\eta$ is assumed to be 
${\eta}_0 \times {\rm MAX}[j/\rho - (j/\rho)_{\rm crit}, 0]$, 
since magnetic reconnection
seems to happen where local $j/\rho$ is larger than some critical
value. 
({\bf B})
The power-spectral densities of light curves above. 
Dotted line is a least square fit with a broken power-law
function for $(j/\rho)_{\rm crit}$ of 200 (open circles). 
It is of great importance to note that general behavior does not depend
on the values of $(j/\rho)_{\rm crit}$.
The unit of time is rotation period at a reference radius 
where the center of initial torus is located. 
The rotation timescale used here will correspond to a few seconds for
BHCs and a few years for AGNs for in realistic situation.}
\end{figure}

The appearance of $1/f^\alpha$ fluctuation, 
or more specifically, the presence of a long-term time correlation
implies a long-distance spatial correlation in the distribution 
of magnetic fields. 
It is thus tempting to examine the spatial magnetic-field distribution.  
We specially pick up the quantity, $j/\rho$,
the ratio of absolute value of electric current density 
to matter density, since it is 
a good indicator regarding a trigger of reconnection (Parker 1994; Ugai 1999). 
In fact, it is shown by MHD simulations that
a fast reconnection, as is observed in solar flares, occurs 
when the electric resistivity becomes anomalously high 
in localized regions (Tajima \& Shibata 1997). 
Such a local, anomalous resistivity can be achieved 
where
electron drift velocity (which is proportional to $j/\rho$) 
exceeds a critical value (Yokoyama \& Shibata 1995).
Hence, any regions with high $j/\rho$ values
are all good candidates for a next reconnection site.

We plot in figure 2 a snapshot of the spatial $j/\rho$ distribution on 
a horizontal plane slightly above the equatorial plane.
The panel roughly demonstrates
to what extent reconnected area expands, 
once reconnection is initiated somewhere.
We notice that the distribution is quite inhomogeneous; 
patchy patterns are visible everywhere in figure 2.  
Importantly, there seems to be no typical size of each patch.
The presence of fractal structure is suspected.
\begin{figure}[t]
\hbox{\psfig{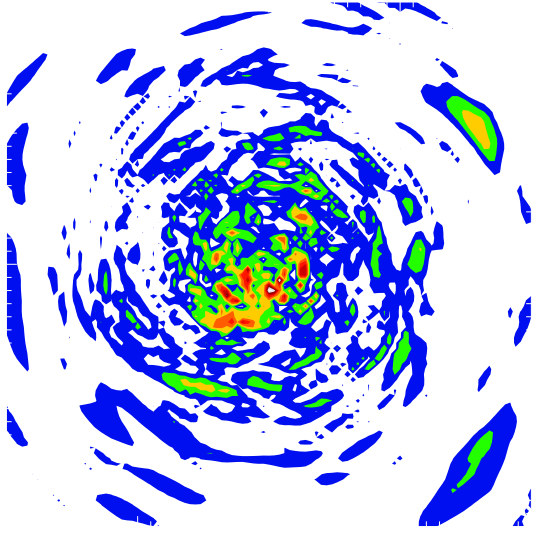}}
\caption
{Color contour map of the $(j/\rho)$ distribution
on a horizontal plane slightly above the equatorial plane.
Here, values of $(j/\rho)$ where colors change are as follows;
70 from white to blue, 120 to green, 200 to yellow, and 300 to red,
respectively.}
\end{figure}

To confirm this idea, we made a fractal analysis for the
three-dimensional MHD disk with mesh point numbers of 
($N_{\rm x}$,$N_{\rm y}$,$N_{\rm z}$) = (100,100,25); namely
we first mark the sites where
$j/\rho$ exceeds some critical value, 
$(j/\rho)_{\rm crit}$, 
and name any assembly of the marked sites clusters.  We then
count the numbers of clusters according to the cluster sizes (i.e.,
volume) for the data sets at five different timesteps and average the counts.
The resultant distribution is plotted in figure 3 with thick lines.
Surprisingly, the cluster size is
distributed in a power-law fashion over three orders of magnitudes of
cluster sizes, from a few larger clusters to numerous smaller clusters.  
\begin{figure}[t]
\hbox{\psfig{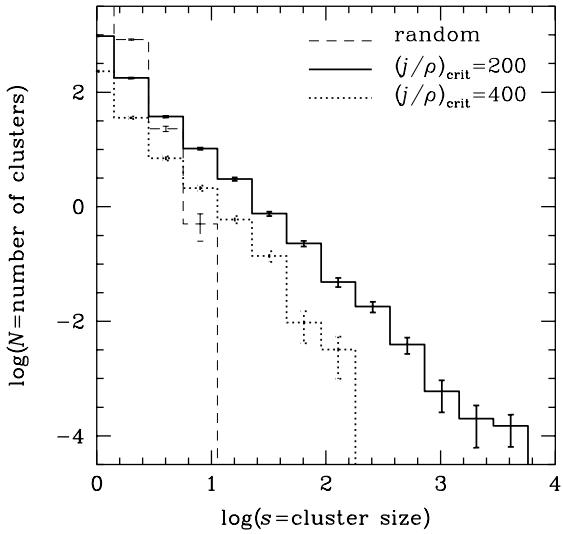}}
\caption
{Histograms of clusters with $j/\rho > (j/\rho)_{\rm crit} = 200$
(by the thick solid lines)
and those with $j/\rho > 400$ 
(by the thick dotted lines) as functions of the cluster size (volume).
Power-law distribution is realized over three orders of cluster sizes;
roughly, $D(s)\propto s^{-2}$.
For comparison, we also plot the same but for
a random distribution (by the thin dashed lines).
Obviously, large clusters are missing.}
\end{figure}

For comparison, we calculated a random distribution in the following way:
we evaluate first how much fraction of the entire volume is covered 
with the marked sites in figure 2, finding about 5.5\% for $j/\rho > 200$.  
We next put a random number between 0 and 1 in each site 
of the three-dimensional box, 
and mark the sites where the number exceeds 0.945.
We then repeat the same procedure done for the MHD disk 
and plot the resultant histogram in figure 3 with the thin dashed line.  
Clearly, there are no very big clusters in the random model.
In other words,
a long-distance $j/\rho$ correlation is lost there.

Magnetic fields in the disk have fractal structure.
To find a fractal dimension, we plot a size (volume) of each cluster 
as a function of its mean radius in figure 4. Here, the cluster mean radius is
defined as $r \equiv \frac{1}{s} {\displaystyle \sum^{s}_{i}}
|{\bf r}_{i}-{\bf r}_{\rm CM}|$ with ${\bf r}_{\rm CM}=\frac{1}{s} 
{\displaystyle \sum^{s}_{i}} {\bf r}_{i}$ for each cluster with a size
of $s (>1)$, where ${\bf r}_i$ is the coordinates of the $i$-th site
belonging to the cluster. 
From the fitting, we find roughly $s\propto r^D$ with $D=1.9$.
\begin{figure}[t]
\hbox{\psfig{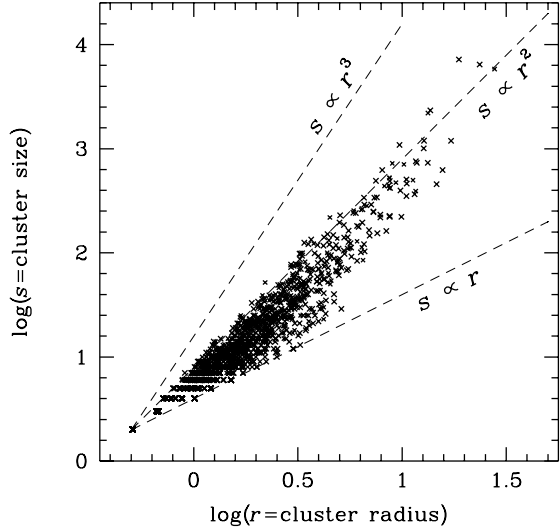}}
\caption
{Relation between the volume ($s$) and mean radius ($r$) of each cluster.
The lines of $s\propto r$, $s\propto r^2$ and $s\propto r^3$ are also 
depicted. The least square fit shows $s \propto r^D$ with $D \sim 1.9$.}
\end{figure}

\section{Discussions}

Then, we address two key questions: what relates the fractal structure to the
temporal fluctuation in the simulated MHD disk?
How can such a fractal distribution arise?
Turbulence, in general, is known to exhibit fractal behavior 
(Procaccia 1984) and 
this fact may be related to our present finding, but predicted fractal
dimension is $D \sim 2.6$, differing from that of the present
case. Alternatively, we 
note the notion of 
self-organized criticality (SOC; Bak 1996; Jensen1998), one of 
the most attractive concepts developed in the study of complex systems.

Bak, Tang \& Wiesenfeld (1988) proposed a sand-pile model to describe
a system exhibiting $1/f^\alpha$ fluctuation.  Suppose that
we fall sand particles one after another on a table.  Fallen
sand particles will form a pile, onto which another sand particle will
be added.  When a slope of the pile in either direction 
exceeds a critical value,
an avalanche occurs and sand particles will slide down in that direction.  
Then, the system spontaneously evolves to and stays at SOC.
In our case, addition of a sand particle corresponds to energy input to
magnetic fields, while the critical slope corresponds to the critical 
$j/\rho$ over which energy dissipation occurs via
reconnection (Mineshige, Takeuchi \& Nishimori 1994).

For systems in a SOC state, long-distance spatial communication
among different sites is naturally built up and it 
yields long-term time correlation.
If each flare light curve is expressed by a time-symmetric
profile with exponential grow and decay about $t=0$, 
$L(t) \propto \exp(-|t|/\tau)$ with $\tau$ ($\leq \tau_{\rm max}$) 
being constant,
its PSD is $P_\tau(f) \propto \epsilon^2/(1+4\pi^2f^2\tau^2)^2$.
If the energy ($\epsilon$) of each flare is distributed as
$N(\epsilon) \propto \epsilon^{-p}$
and each flare duration is related to energy as, $\epsilon \propto \tau^D$,
the total PSD becomes
\begin{eqnarray*}
   P(f) &=& \sum^{{\tau}_{\rm max}}_\tau P_\tau(f)N(\tau)\Delta \tau \\
      &\propto& \sum^{{\tau}_{\rm max}}_\tau
                 \frac{\tau^{2D}}{(1+4\pi^2f^2\tau^2)^2} (\tau^D)^{-p} 
                 \frac{d\epsilon}{d\tau}\Delta \tau ,
\end{eqnarray*}
\begin{equation}
  \quad \longrightarrow
   P(f) \propto \left(\frac{1}{f}\right)^{(3-p)D} F(f) \qquad 
   (\Delta \tau \rightarrow 0).
\end{equation}
Here, $F(f)$ 
[$=\int^{2\pi \tau_{\rm max} f}_0 dx (x^{(3-p)D-1})/(1+x^2)^2$] is a 
slowly varying function of $f$ for $f > 1/(2 \pi \tau_{\rm max})$.
We may regard $\epsilon \propto s$ (volume of a clump) 
and $\tau \propto r$ (mean radius of a clump). 
Then,
we find from the simulation that $p \sim 2$ and $D \simeq 2$ (see figure 4).
Hence equation (1) leads $P(f) \propto f^{-2}$,
in agreement with the numerical result (see figure 1).
That is the reason why the fractal magnetic field 
produces $1/f^\alpha$ fluctuations 
(Takeuchi, Mineshige \& Negoro 1995; Kawaguchi et al.\ 1998).

One of the most conspicuous natures of the SOC is its ubiquity;
namely, it is supposed to describe various non-equilibrium open systems,
such as earthquakes, forest fires, evolution of biological species,
and traffic flows (Bak 1996).
In astrophysical context, it is important to note that
coronal magnetic fields in the Sun are suggested to be in a SOC
state (Lu \& Hamilton 1991; Vassiliadis et al.\ 1998) as well, 
thus exhibiting power-law occurrence rate of flares 
and $1/f^\alpha$ fluctuations in solar flare curves (UeNo et al.\ 1997).  

Gamma-ray bursts (GRBs) also occasionally exhibit 
$1/f^\alpha$ fluctuations (Beloborodov, Stern \& Svensson 1998).
In some models of GRBs which involve merger
of two compact objects (white dwarf, neutron star, and black hole)
a sort of an accretion disk is thought to be formed by debris of 
one component around the other (M\'{e}sz\'{a}ros 1999).  
The situation could be similar. 
We thus expect
frequent reconnection events with a smooth size (amplitude and duration) 
distribution to occur in GRBs, 
which will give rise to $1/f^\alpha$ fluctuations 
(Panaitescu, Spada \& M\'{e}sz\'{a}ros 1999).
Likewise, any other magnetic systems, regardless of system size,
may show similar effects.

\par
\vspace{1pc} \par
The authors would like to thank Prof.\ S.Kato and
Prof.\ T.Tajima for their useful comments.
This work was supported in part 
by Research Fellowships of the Japan Society for the
Promotion of Science for Young Scientists (4616, TK), and
by the Grants-in Aid of the
Ministry of Education, Science, Sports, and Culture of Japan
(10640228, SM).
This work was partially supported by 
Japan Science and Technology Corporation.
Numerical computations were carried out by using
Fujitsu VPP300/16R at National Astronomical Observatory, Japan.

\section*{References}
\small
\re
Abramowicz, M. A., Lanza, A., Spiegel, E. A., \& Szuskiewicz, E. 1992,
 Nature, 356, 41
\re
Bak, P., Tang, C., \& Wiesenfeld, K.\ 1988, Phys. Rev. A, 38, 364
\re
Bak, P.\ 1996, How Nature Works (Springer-Verlag, New York)
\re
Beloborodov, A.M., Stern, B.E., \& Svensson R.\ 1998, ApJ, 508, L25
\re
Cable S., \& Tajima T.\ 1997, Comm.\ Plasma Phys.\ Controlled Fusion, 
18, 145
\re
Galeev, A.A., Rosner, R., \& Vaiana, G.S.\ 1979, ApJ, 229, 318
\re
Jensen, H.J.\ 1998, Self-Organized Criticality 
 (Cambridge Univ. Press, Cambridge)
\re
Kato, S., Fukue, J., \& Mineshige, S.\ 1998, Black-Hole Accretion Disks
 (Kyoto Univ. Press, Kyoto)
\re
Kawaguchi, T., Mineshige, S., Umemura, M., \& Turner, E.L.\ 
 1998, ApJ, 504, 671
\re
Lu, E.T., \& Hamilton, R.J.\ 1991, ApJ, 380, L89
\re
Machida, M., Hayashi, M. \& Matsumoto R.\ 1999, submitted to ApJL
\re
M\'{e}sz\'{a}ros, P.\ 1999, To appear in Black
 Holes and Gravitational Waves. (ed Nakamura, T., Supplement of
 Prog.\ Theor.\ Phys)
\re
Mineshige, S., Takeuchi, M., \& Nishimori, H.\ 1994, ApJ, 435, L125
\re
Mineshige, S., Kusunose, M., \& Matsumoto, R.\ 1995, ApJ, 445, L43
\re
Narayan, R., McClintock, J. E., \& Yi, I.\ 1996, ApJ, 457, 821
\re
Negoro, H., Kitamoto, S., Takeuchi, M., \& Mineshige, S.\ 1995,
ApJ, 452, L49
\re
Panaitescu, A., Spada, M., \& M\'{e}sz\'{a}ros, P.\ 1999,
 submitted to ApJ, astro-ph/9905026
\re
Parker, E. N.\ 1994, Spontaneous Current Sheets in Magnetic Fields 
 (Oxford Univ. Press, New York) p50
\re
Press, W.H.\ 1978, Comm.\ Astrophys., 7, 103
\re
Procaccia, I.\ 1984, Journal of Statistical Physics, 36, 649
\re
Shibata, K.\ 1996, Adv.\ Space Res., 17, (4/5)9
\re
Tajima, T., \& Shibata, K.\ 1997, Plasma Astrophysics (Addison
Wesley, Massachusetts, ) ch2.6
\re
Tajima, T., \& Shibata, K.\ 1997, Plasma Astrophysics (Addison
Wesley, Massachusetts) p237
\re
Takeuchi, M., Mineshige, S., \& Negoro, H.\ 1995, PASJ, 47, 617
\re
UeNo, S.\ et al.\ 1997, ApJ, 484, 920
\re
Ugai, M.\ 1999, Phys. Plasma, 6, 1522
\re
Ulrich, M.-H., Maraschi, L., Urry, C.M.\ 1997, ARA\&A, 35, 445
\re
van der Klis, M.\ 1995, in X-Ray Binaries, 
ed W.H.G. Lewin et al. (Cambridge University Press, Cambridge) p252 
\re
Vassiliadis, D., Anastasiadis, A., Georgoulis, M., \& Vlahos, L.\ 1998, 
ApJ, 509, L53
\re
Wheeler, J.C.\ 1977, ApJ, 214, 560
\re
Yokoyama, T. \& Shibata, K.\ 1995, Nature, 375, 42
\re

\label{last}

\end{document}